\documentstyle[preprint,aps,epsfig]{revtex}
\begin{document}
\draft
\tighten
\preprint{JLAB-THY-01-12}
\title{
Extracting forward strong amplitudes from elastic differential cross
sections}
\author{C. M. Chen}
\address{Saint John's and Saint Mary's Institute of Technology,
Tam-Sui\\Taipei County, Taiwan 25135, ROC}
\author{D.~J.~Ernst}
\address{Department of Physics and Astronomy,
Vanderbilt University,
Nashville, TN  37235 \\
Jefferson Lab, 12000 Jefferson Avenue, Newport News, VA 23606}
\author{Mikkel B.~Johnson}
\address{Los Alamos National Laboratory,
Los Alamos, NM  87545}
\date{\today}
\maketitle
\begin{abstract}
The feasibility of a model-independent extraction of the forward strong
amplitude from elastic nuclear cross section data in the Coulomb-nuclear
interference region is assessed for $\pi$ and $K^+$ scattering at
intermediate
energies. Theoretically-generated ``data" are analyzed to provide criteria
for
optimally designing experiments to measure these amplitudes, whose energy
dependence (particularly that of the real parts) is needed for disentangling
various sources of medium modifications of the projectile-nucleon
interaction.
The issues considered include determining the angular region over which to
make
the measurements, the role of the most forward angles measured, and the
effects of statistical and systematic errors.  We find that 
there is a region near the forward direction where Coulomb-nuclear
interference
allows reliable extraction of the strong forward amplitude for both pions
and the $K^+$ from .3  to 1 GeV/c.
\end{abstract}

\section{INTRODUCTION}

In the scattering of a charged particle from a nucleus, the measured elastic
differential cross section is the square of a scattering amplitude,
$F_{el}(\theta)$, which is
the sum of a Coulomb amplitude $f_c(\theta)$ and a strong amplitude
$F_N(\theta)$,
\begin{equation}
     F_{el}(\theta)=f_{c}(\theta)+F_{N}(\theta)\,\,,
\label{eq:0}
\end{equation}
Many outstanding issues of strong interaction physics may be addressed if
one
can determine $F_N(\theta)$ from the experimental data.  For example,
for pion and kaon beams, medium modifications of the underlying 
projectile-nucleon scattering amplitude are of great current
interest~\cite{sie84,che95}, and efforts are being made
to extract this information from scattering data.  Typically,
modern analyses utilize sophisticated optical model 
codes~\cite{che95,gie88}
based on microscopic models of the underlying dynamics, in which direct
comparisons are made between the differential cross sections calculated by
the codes and measured differential cross sections.

We wish to emphasize that for pion-nucleus scattering,
especially in the GeV range of energies and for kaon-nucleus scattering at
energies above several hundred MeV, it is extremely valuable for the 
theorists 
to have precise information on the $\it amplitude$ $F_N(\theta)$ for $\theta 
\approx 0
^\circ$.
Measurements of $d\sigma_{el}/d\Omega(\theta)$ at small $\theta$ contain
this
information through the interference of the Coulomb and strong interactions
in Eq.~\ref{eq:0}.  Determinations of Re$F_N(0)$ and Im$F_N(0)$ are strong
constraints on the underlying models.  We make the case in this paper
that the considerable inconsistency existing in currently available data 
sets~\cite{che98}
can be remedied at existing facilities with appropriate measurement 
strategies.  We hope that our results will facilitate precise measurements
of $d\sigma_{el}/d\Omega(\theta)$ at small $\theta$ at these laboratories.

The issues involved in designing measurement strategies may be examined
through application of optical models such as those mentioned above.
However,
a much simpler approach is possible due to the fact that~\cite{che98} for
$\pi$-nucleus scattering at energies above the $\Delta_{33}$ resonance
($T_\pi\ge 300$ MeV) and K$^+$-nucleus scattering at all energies, the
region over which the Coulomb amplitude is an appreciable fraction of the
strong amplitude extends from a few degrees out to
angles generally beyond ninety degrees. This is an ideal situation for
utilizing Coulomb-nuclear interference in a model independent way to extract
the strong amplitude from elastic differential cross section measurements.

To accomplish this, elastic differential cross section data needs to be
taken
in the near forward direction.  Such measurements do not require a large
amount
of beam time; the measurements are all made at angles where the differential
cross section is quite large. A set of data spanning a large energy region
at
reasonably spaced energy intervals is thus possible. Data taken with
$P_{lab}$
at intervals of 25 MeV/c from 300 MeV/c to 1 GeV/c would take less time than
is
required for the measurement of a single differential cross section out to a
reasonable angle. This is because the differential cross section is
diffractive
and thus falls exponentially with angle. Points with decent statistics at
large
angles where the cross section has typically fallen by four orders of
magnitude
require more beam time than do a large number of measurements at small
angles.

The primary purpose of the present work is to assess the feasibility of
measuring the real and imaginary parts of the forward scattering amplitude
with sufficient accuracy to clarify the issues alluded to above.  Our
secondary
purpose is to present enough of the details of the argument and analysis so
that the experimentalists can optimally design such an experiment.

In Sect. II, we review a simple procedure for extracting the strong amplitude 
from the elastic differential cross section and provide a summary of the 
results of our analysis (details are given in the Appendices) based on this 
method.  We utilize the model of 
Refs.~\cite{che95,gie88} to generate theoretical elastic differential cross 
sections at a set of angles. These results constitute model data to which we 
add statistical or systematic errors of various magnitudes. The momentum-space 
model of \cite{che95,gie88} produces results which are very similar to the 
measured elastic differential cross sections for $\pi$ \cite{tak95,mar90} and 
for $K^+$ \cite{mar90,mic96}. We thus believe that the analysis done here on 
these model data sets is directly applicable to real data. In Section III  
we discuss underlying physics issues involved. The final section 
summarizes the paper and presents our overall conclusions.

\section{PROCEDURE AND RESULTS}

To use Coulomb-nuclear interference to extract the forward strong 
amplitude from the elastic differential cross section, we envision making an 
expansion of $F_N(\theta )$ and fitting the coefficients of 
the various terms to data.  In order to determine this amplitude most 
effectively, certain questions need to be addressed.  These would include 
the following:  what is the  optimal angular range over which measurements
should be made; to what order should one expand the strong 
amplitude; how important is it to measure to the most forward-possible angle;
how do statistical and systematic errors affect the extracted values of the 
strong forward amplitude; and, most importantly, how stable and reliable is
the method for extracting the forward strong amplitude?  We define the 
procedure we use for answering these questions in Sect.~II.A and present our
results in Sect.~II.B, below.

\subsection{Procedure}

The nuclear amplitude $F_{N}(\theta)$ that we will use for our analysis
is defined in terms of the ${\it point}$ Coulomb scattering amplitude
$f_{c,pt}(\theta)$,
\begin{equation}
     F_{el}(\theta)=f_{c,pt}(\theta)+F_{N}(\theta)\,\,.
\label{eq:1}
\end{equation}
Alternatively,
we could define the nuclear amplitude relative to the extended Coulomb
amplitude or introduce the strong amplitude related to $F_N(\theta)$
through the Bethe phase~\cite{coo77,bet58,wes68}.  All such definitions
are mathematically equivalent, and we choose to work with the definition
(the one corresponding to Eq.~\ref{eq:1}) that simplifies the empirical 
analysis.
Since we anticipate the use of this amplitude as a constraint on the optical
model, all three nuclear amplitudes are equally suitable for this purpose.

The simplest procedure for obtaining $F_N(0)$ is based on a Taylor series 
expansion of $F_N(\theta)$ in powers of $\sin^{2}\theta/2$.  Truncating the 
series after three terms, we 
then have six real parameters, $A_{R}$, $B_{R}$, $C_R$, $A_{I}$, $B_{I}$, and 
$C_I$ defined by:

\begin{equation}
F_{N}(\theta)=A_{R}\left(1-B_{R}\sin^{2}\theta/2-C_R\sin^4\theta/2\right)
     +i\, A_{I}\left(1-B_{I}\sin^{2}\theta/2-C_I\sin^4\theta/2\right)\,\,.
\label{eq:2}
\end{equation}
The experimental differential cross section minus the point Coulomb
differential cross section is given by $d\sigma_N / d\theta$
\begin{eqnarray}
{d \sigma_N \over d\theta}(\theta)&\equiv&{d\sigma_{el} \over d\theta}
(\theta)-
{d\sigma_{c,pt}\over d\theta}(\theta)\nonumber\\
&=&\vert\,F_N(\theta)\,\vert^2+ 2\,{\rm
Re}\,\left[ f_{c,pt}(\theta)\,F_N^*(\theta)
\right]\,\,,
\label{eq:3}
\end{eqnarray}
where $d\sigma_{el} / d\theta$ is the experimental elastic differential
cross section and $d\sigma_{c,pt} / d\theta$ is the point Coulomb differential
cross section.  The various expansion parameters of $F_N(\theta)$ given in 
Eq.~\ref{eq:2} may then be determined by a fit of Eq.~\ref{eq:3} to the
forward angle experimentally measured  differential cross sections,
$d\sigma_{el}/ d \theta$. 

This work assesses the above procedure for extracting the forward strong 
amplitude based on theoretically generated
model data sets. We utilize the  the momentum-space theory of
Ref.~\cite{che95,gie88} to generate the model  data. To study the effects of
statistical errors, Gaussian distributed errors are added to the
theoretically
generated model data. To understand the effects of systematic errors, these
data sets are renormalized both upward and downward. We do the analysis for
both $\pi^-$ and $K^+$ at $P_{lab}=500$ Mev/c and 1 GeV/c. This study 
demonstrates that the extraction of the forward strong amplitude, both the 
real and imaginary part, is feasible, and we
provide the information needed to optimally design  an experiment.

The strong interaction amplitude extrapolated to zero degrees is then given by  
$F_{N}(0)=A_{R}+i\, A_{I}$.  As we have stated, once $F_N(0)$ is obtained, it 
may be used in conjunction with optical model descriptions such as those of 
Ref.~\cite{che95,gie88} to constrain the underlying microscopic models.  If 
desired, the real and imaginary parts of the strong amplitude may be extracted 
directly \cite{coo77,bet58,wes68} from $A_R$ and $A_I$ using a Bethe-phase
analysis~\cite{che98}.  Note that the real part
of the forward strong amplitude can also be extracted from transmission
experiments \cite{coo76}, a technique that has been applied \cite{jep83} to
resonance energy $\pi$-nucleus scattering.

\subsection{Results}

The details of the assessment of the above procedure for obtaining $F_N(0)$ 
are provided in the Appendices.  In 
order to determine the feasibility of an experiment, we examine the angular 
range over which data should be taken (Appendix 1), the effects of statistical 
(Appendix 2) and systematic (Appendix 3) errors, and the importance of taking 
very forward angle points (Appendix 4). These details should prove 
valuable for the design of an experiment. Here we summarize our results. Our
goal is to extract the forward amplitude, i.e. 
$A_I$ and $A_R$. In the Appendices we also provide some guidance on the 
possibility of extracting the next term in the expansion, $B_I$ and $B_R$.

We find in Appendix 1 that there is a substantial angular region over which 
the strong and Coulomb amplitudes are reasonably comparable. We also find that 
there is an angular region where the strong amplitude is linear in 
$\sin^2\theta/2$. We explored going beyond linear order in the expansion 
by extending the angular region of the measurements. This extension 
did {\it not} improve the 
ability to extract the forward amplitude. The smallest angle of this region
determined by requiring the magnitude of the Coulomb amplitude to be twice the 
size of the strong amplitude. This implies, for both pions and $K^+$, that 
$\theta_{min}=4^\circ$ for 500 MeV/c and $\theta_{min}=2^\circ$ for 1 GeV/c. 
The maximum angle is determined by the conditions that the linear 
approximation hold and that the extrapolation yield an accuracy to, say, 
better than 1\%. This would require a measurement out to about $16^\circ$. 

In Appendix 2 we find that the magnitude of the error in the extracted forward 
amplitude is 
proportional to the statistical errors in the data. For $\pi^-$ the percent 
error in $A_I$ is approximately equal to the percent error in the data. The 
error in $A_R$ is of the same magnitude, but, because the value of $A_R$ is 
much smaller, it corresponds to larger percent error in this quantity.
The situation is 
different for $K^+$. The percent error in $A_R$ is approximately three times 
the error in the data at both energies examined, while for $A_I$ it is seven 
times at 500 MeV/c and three times at 1 GeV/c.

Because the cross sections are large throughout the angular region for these 
experiments, it is easy to have the statistical errors much smaller than the 
systematic errors.  For systematic errors above about 10\%, the analysis 
becomes nonlinear and is not reliable, as discussed in Appendix 3.
Systematic errors of 5\% or less do yield reliable results. 
For $\pi^-$ scattering at 500 MeV/c, each 1\% systematic 
error yields only 0.3\% error in the extrapolation for $A_R$ and 0.5\% for 
$A_I$. At 1 GeV/c, the behavior of $A_I$ is nonlinear, but stable. There is 
never more than 0.7\% error in the extracted value for each 1\% systematic 
error. The error in $A_R$ is comparable in magnitude to the error in $A_I$,
which, again, corresponds to a much larger percent error in $A_R$ at
1 GeV/c.  For $K^+$ 
each 1\% systematic error produces at 500 MeV/c an error of 5.7\% for $A_I$, 
1.5\% for $A_R$, and at 1 GeV/c an error of 1.4\% for $A_I$ and 3\% for 
$A_R$. Systematic errors below 5\% would very significantly constrain 
theories for both pions and $K^+$.

Since the normalization of the incident beam can be a large source of 
systematic error, a special setup that would allow for measurements 
into the far forward direction would be desirable.  Measurements at the very 
forward angles where the Coulomb scattering becomes dominant would allow the 
normalization of the beam to be determined by comparing the measurements to 
the known Coulomb amplitude.

Finally, the forward-most points are the most difficult to take and yet we 
would expect them to be the most important in controlling the extrapolation to 
$\theta=0^\circ$. We examine in Appendix 4 the consequences of increasing the 
errors on the first two experimental points at $\theta_{min}$ and 
$\theta_{min}+0.5^\circ$, or eliminating them altogether.  For statistical 
errors of 1\% or 2\%, the removal of these data points roughly doubles the 
magnitude of the errors in the extracted quantities, with the exception of 
$A_R$ for the $K^+$ where at 500 MeV/c the extracted value is little affected 
and at 1 GeV/c  it increases by a factor of three. The forward points taken 
with good statistics and with systematic errors no worse than the remaining 
points can, in general, reduce errors by a factor of two or more.

\section{Discussion}

The strong amplitude at zero degrees as a function of energy contains
information which is complimentary to that obtained by measuring exclusive
cross sections as a function of angle. The imaginary part of the forward
strong
amplitude is related through unitarity to the total cross section. Its
extraction from differential cross section measurements is an experimental
check on the consistency of two independent experimental measurements,
elastic
scattering and transmission experiments. Because it is independent of the
total cross section, the real part of the forward strong amplitude provides 
new information to help decide among competing models.  Moreover, the real 
part of the strong amplitude at zero degrees is a quantity that is not driven 
by small corrections to the theory, such as is the exact depth of the minima 
in the diffraction
pattern. It is a qualitative feature of the reaction, which like the total
cross section, puts constraints on the theory at the qualitative level.

There is a growing body of evidence \cite{sie84} that the kaon
interaction with a nucleon is enhanced in the nuclear medium 
compared to that in free space. Several phenomenological analyses
\cite{che98b}
of this  data have been performed, but a consistent picture of the
underlying
physics has not yet emerged. Measurement of the strong forward scattering
amplitude for kaon-nucleus elastic scattering as a function of energy would
provide an independent check on the total cross section measurements. 
Since the real part of the forward amplitude contains new information, it 
would presumably help determine the underlying physics behind the in-medium 
increase in the interaction.

The situation is somewhat different for the pion. The $\pi$-nucleon
interaction is dominated by a number of overlapping resonances. Evidence from
photo-reactions indicate the existence of an in-medium 
modification \cite{kon94} of the resonances (their mass, width, and coupling 
constants). In Ref.~\cite{che95}, modifications of the properties
of the excited hadrons were taken from the photo-reaction \cite{kon94} and
the effects on $\pi$-nucleus total reaction cross sections were predicted
and compared with data from \cite{clo74}. From that work (see also 
\cite{ari95}) one would conclude that there is a medium enhanced two-body 
cross section for the $\pi$ similar to that found for the 
$K^+$. However, because of the inconsistencies referred to above~\cite{che98},
this result is not as convincing as one would like.  Data of the type that we 
discuss here, namely precise $\pi$-nucleus elastic scattering 
in the forward direction from 300 MeV/c to 1 GeV/c, could resolve the existing 
discrepancy and provide a critical quantitative characterization of the
medium effect for pions.  

The data would provide, at the same time, the real part of the forward strong 
amplitude as a new and important clue to the underlying dynamics.  For the 
$\pi^-$, models suggest that $F_N(0)$ may pass through zero just below 
1 GeV/c~\cite{che98}.  This situation presents an interesting physics 
opportunity arising
from the fact that this zero is strongly associated with a zero in the 
real part of a two-body amplitude dominated by numerous resonances 
centered at various energies throughout the GeV region.  The signs and 
magnitudes of the real 
parts of the corresponding partial waves contributing to the amplitude 
occur 
in such a way that their sum vanishes close to 1 GeV/c.  The precise energy at 
which this amplitude crosses zero for a nuclear target would thus be a 
sensitive measure of medium-induced mass shifts for this set of resonances.

\section{Summary  and Conclusions}

We find that for intermediate energy $\pi$ and $K^+$ elastic scattering
there
is a region near the forward direction where Coulomb-nuclear
interference
allows reliable extraction of the strong forward amplitude. We find that
statistical and systematic errors in the data are reflected in an
understandable and predictable way in the errors in the extracted value of
$F_N(0)$, as long as the errors are kept at less than about 5\%. We provide
guidance for the angular region over which to take the data. The accuracy of 
the extrapolation to zero 
degrees is sensitive to how far forward one can take data, as is demonstrated 
in Tables II and III of Appendix 2.  In addition, 
taking data to even smaller angles where 
the known Coulomb interaction dominates could be a way to control the 
systematic error associated with beam normalization.

New data making use of such an analysis could resolve limitations of and
outstanding disagreements among various data sets that exist.  For example,
in Ref.~\cite{che98} the elastic differential cross section data
\cite{tak95}
for $\pi^-$ scattering from $^{12}$C at $P_{lab}=610$, 710, 790, and 895
MeV/c
was used to extract the zero-degree elastic scattering amplitude.
The total cross sections determined from the imaginary
part of the zero-degree amplitude were consistently lower than 
those \cite{clo74} measured by transmission experiments. Additionally, the 
real part of the zero degree scattering amplitude was found to
have an energy dependence that is not present in theoretical calculations
\cite{gie88}. There exist other elastic differential cross section
measurements for pions \cite{mar82} but the quality of these data do not
allow
a stable extraction of the forward strong scattering amplitude.
The data \cite{mar90,mic96} on elastic $K^+$--nucleus scattering is even more
limited. Although the results produced total cross sections
that were consistent with transmission measurements~\cite{kra92}, in
order to extract the forward strong amplitude from the data of \cite{mic96},
model assumptions had to be made.

We have provided criteria that will enable the optimal design of an
experiment.
Since the imaginary part of $F_N(0)$ is related by unitarity to the total
cross
section, this quantity will provide an independent check on its measured
values. The real part of $F_N(0)$ is a quantity which would offer
considerable
constraint on theories. It would provide information on the important
question
of the medium modification of the properties of both the target nucleons
and,
in the case of $\pi$ scattering, the produced excited nucleons. For $\pi$
scattering, the energy at which the real part of the amplitude passes
through
zero would be a good measure of medium-induced mass shifts of the excited
hadrons. In this case data for both $\pi^+$ and $\pi^-$ would be valuable; a
consistent theoretical treatment of these two cases would indicate that all
of
the Coulomb effects had been adequately accounted for.

A convincing understanding of the underlying physics that is determining
intermediate energy meson-nucleus reactions will require a consistent
interpretation of a number of reactions. In addition to elastic scattering
and total cross sections that are the subject of this work, this would
include
quasielastic scattering of kaons \cite{kor93} and pions, both with
\cite{pet92}
and without \cite{wis93} charge exchange. The measurement of Re $F_N(0)$
would
be an important ingredient in this broader program.

\acknowledgements The work of DJE is supported, in part, by the
U.S. Department of Energy under grant DE-FG02-96ER40975. The Southeast 
Universities Research Association (SURA) operates the Thomas Jefferson 
National Accelerator Facility under DOE contract DE-AC05-84ER40150. The work 
of MBJ was supported, in part, by the US Department of Energy under contract 
No.~W-7405-ENG-36. The work of CMC was supported, in part, by the ROC 
National Science Council under grant NSC 86-2112-M-129-001. CMC and DJE would 
like to thank the Los Alamos National Laboratory for their hospitality during 
part of this work.

\appendix
\section*{}

\subsection{Angular range}

The first question is to determine the angular region over which 
$f_{c,pt}(\theta)$ and $F_N(\theta)$ are comparable in size. In
Figs.~\ref{fig1}-\ref{fig4}
we plot the theoretical values for the real and imaginary parts of
$f_{c,pt}(\theta)$ and  $F_N(\theta)$ versus $k^2\sin^2\theta/2$ for elastic
scattering of $\pi^-$ and $K^+$ from $^{12}$C at $P_{lab}=$ 500 MeV/c and 1
GeV/c. First, we see that in general the amplitudes are non-zero and of
comparable size for at least a part of the angular region depicted. The
exception to this is Re  $F_N(\theta )$ for $\pi^-$ at 1 Gev/c. For the
pion,
Re $F_N(0)$ passes  through zero at a momentum just below 1GeV/c and is thus
small and not of a typical size at  1 GeV/c. One also notices that Im
$f_{c,pt}(\theta )$ goes to zero for  increasing $\theta$ as expected. From
Eq.~\ref{eq:3} we see that the measured cross section is still dependent on
both the real and imaginary parts of $F_N(\theta )$ and thus our methodology
is
still valid even though the Coulomb amplitude has become nearly real. We
will
find that it is the convergence of the Taylor series that  determines the
maximum angle $\theta_{max}$ out to which data need be taken. We also see
that
there is a significant range over which the strong amplitude is nearly
linear
in $\sin^2\theta/2$. If the goal is to extract $A_R$ and $A_I$, then  a
linear
expansion which includes only $A$'s and $B$'s should suffice.

For the minimum angle $\theta_{min}$ at which to take data, we adopt the
criteria that the point Coulomb amplitude $f_{c,pt}(\theta_{min})$ be
approximately one half of the strong amplitude $F_N(\theta_{min})$. Points
where the Coulomb amplitude is dominant will carry little information about
the
strong amplitude and thus should not be included in the analysis. From
Figs.~\ref{fig1}-\ref{fig4}, we find that for both pions and $K^+$ this
implies
$\theta_{min}= 4^\circ$ at 500 MeV/c and $\theta_{min}= 2^\circ$ at 1 GeV/c.
Because the physical size of a spectrometer can limit how far forward
measurements can be made, we revisit in Appendix 4 the question of
how important are the small angle data points for determining $F_N(0)$.

Given $\theta_{min}$, what is the optimal value for the maximum angle
$\theta_{max}$ at which to take data? The answer to this question will
depend
on the accuracy to which one wishes to work. Our point of view will be that
we
wish to learn both the real and the imaginary part of $F_N(0)$, i.e. $A_I$
and
$A_R$. To find $\theta_{max}$, we generate theoretical values of $d\sigma_N
/d\theta$ at a discrete set of points $\theta_i$. We then fit \cite{MIN} the
polynomial expansion for $F_N(\theta_i)$ to determine the expansion
coefficients. As we increase the number of points, and hence the value of
$\theta_{max}$, the values of $A_R$ and $A_I$ from the fit will increasingly
differ from the exact values. The results of these calculations are given in
Figs.~\ref{fig5}-\ref{fig8} for $\pi^-$ elastic scattering from $^{12}$C at
500
Mev/c and 1 GeV/c, and similarly in Figs.~\ref{fig9}-\ref{fig12} for $K^+$.
The
curves were generated with $\theta_i$ starting at $\theta_{min}$ and then
taking evenly spaced points at intervals given by
$\delta\,\theta=0.5^\circ$.
In each figure the solid line represents the results of using two terms,
i.e.
the $A$'s and $B$'s, in the Taylor series; the dashed lines use three terms.
Given a desired accuracy for $A_R$ or $A_I$, the value of $\theta_{max}$ can
be
determined from these graphs. For example, we have included on the graphs as
dotted lines the exact value of $A\pm$ 1\%. Using two terms in the expansion
and insuring that the value of $A_R$ is valid to better than 1\%,
Fig.~\ref{fig5} suggests that at 500 MeV/c one must use fewer than 23 points.
Here 23 points corresponds to
$\theta_{max}=\theta_{min}+23\times\delta\,\theta=4^\circ +23\times
0.5^\circ
=  15.5^\circ$. Utilizing a three-term expansion, this becomes 48 points
($\theta_{max}= 28^\circ$). Looking at Fig.~\ref{fig6} one finds that 
for $A_I$ at 500 MeV/c, a 1\% error gives 22 points ($\theta_{max}=
16^\circ$) for a two term expansion and  45 points ($\theta_{max}=
26.5^\circ$)
for the three-term expansion.

We have found that in fitting to the model data, even in the case of no
statistical errors, there are multiple local minima in $\chi^2$ as a
function
of the expansion coefficients. However, for the cases we examined in this
work,
the absolute minimum always corresponded to parameters which go continuously
to
the exact answer as the model data was improved. Thus for the level of
errors
we have investigated the technique is stable.

\subsection{Effect of statistical errors}

To understand the role statistical errors in the data have in generating
errors
in the extracted value of $F_N(0)$, we generate model data by adding
Gaussian
distributed random errors to each of the values of the cross section
$d\sigma
/  d\theta(\sigma_i)$. We do this ten times to generate ten model data sets.
We
utilize a two term expansion and thus extract $A_R$, $A_I$, $B_R$, and $B_I$
for each of these data sets. From the ten sets we can get the average of
each
extracted parameter and its standard deviation. We do this entire process
three
times, setting the errors in the model data to produce standard deviations
of
1\%, 2\%, and 5\%.

The results are presented in Table \ref{table1}, where the
results are given for $\pi^-$ and $K^+$ elastic scattering from $^{12}$C at
$P_{lab}=500$ MeV/c and 1 GeV/c. From this table, some general
guidelines can be determined concerning the errors in the extracted expansion
coefficients as a function of the errors in the data. For $\pi^-$ the
percent
error in $A_I$ is approximately equal to the percent error in the data. This
is
not surprising as $A_I$ is the dominant term in the expansion. The error for
$A_R$ is roughly equal in absolute magnitude to the error for $A_I$. Since
$A_R$ is smaller than $A_I$ the percent error in $A_R$ is larger. One might
also wish to reliably extract the $B$ coefficients. The percent error in the
$B_I$ coefficient is approximately five times larger than the percent error
in
the data. The $B_R$ coefficient is poorly determined; it would take an
exceptionally precise experiment to learn anything about it. Adding in the
$C$
coefficients to the expansion and including data to larger angles was found
not
to improve the situation. We do not recommend working with a three-term
expansion.

For $K^+$ elastic scattering the situation is somewhat different. The values
of
$A_R$ and $A_I$ are comparable.  The percent error in $A_R$ is approximately
three times the percent error in the data; the percent error in $A_I$ is
approximately seven times the percent error in the data at 500 MeV/c and
three
times at 1 GeV/c. As was the case for $\pi^-$, it does not appear possible
to
extract a reliable value of $B_R$. At $P_{lab}=500$ MeV/c,  the value of
$B_I$
can be determined only at the level of its magnitude and sign, while at 1
GeV/c
it can be determined at the level of a percent error that is ten times the
percent error of the data.

\subsection{Effect of systematic errors}

Since the data is to be taken in the forward direction where count rates are
high, it should be possible to have the statistical errors smaller than the
systematic errors, particularly if good quality data can be taken at
$\theta_{min}$. To study the effects of systematic errors, we take the model
data and increase it  uniformly by 5\% and by 10\%. We have also decreased
the
model data uniformly  by 5\% and 10\% and find results that generally
indicate
a reasonably linear effect.

In Table \ref{table4} we present results for $\pi^-$  and $K^+$ elastic
scattering from $^{12}$C at $P_{lab}= 500$ MeV/c and 1 Gev/c when we
increase
and decrease the model data by 5\% and 10\%. The first thing to note is that
if the systematic errors are too large the error in the extracted
coefficients
can become a nonlinear function of the size of the systematic error. This
occurs here for $+10$\% systematic error and $\pi^-$ and $K^+$ scattering at
500 MeV/c  as can be seen in the behavior of $A_I$. Experiments should thus
keep the systematic errors at 5\% or less to avoid this possibility.

For $\pi^-$ scattering at 500 MeV/c, the value of $A$ is particularly
stable;
for every 1\%  systematic error in the measured cross section, there  is
only
0.3\% error in the extracted value of $A_I$ and 0.5\% error in the
extracted
value of $A_R$. The errors in the extracted values of $A_I$ for $\pi^-$ at 1
GeV/c  are not linear. However, the sign of the nonlinearity is such that
it leads to a stable  value for $A_I$. For each 1\% systematic error there is 
never more
than a 0.7\% error  in the extracted value of $A_I$. As noted earlier for
$\pi^-$ at 1 GeV/c, the value  of $A_R$ is atypically small thus giving
errors
which on a percentage basis are large. There is a 14\% error in the
extracted
value of $A_R$ for each 1\% systematic error.  If one takes a typical value
for
$A_R$ over this energy region of 3.0 fm, then one finds that the error in
$A_R$
is  2\% for every 1\% systematic error.

For $K^+$ scattering at 500 MeV/c we find that every 1\% of systematic error
produces 5.7\% error in the extracted value of $A_I$ and 1.5\% in the
extracted
error of $A_R$. At 1 GeV/c these become 1.4\% for $A_I$ and 3\% for $A_R$.
For
$K^+$ the value of $A_R$ can more accurately be determined at 500 MeV/c and
this gradually changes as the momentum increases so that $A_I$ is more
accurately determined at 1 GeV/c.

\subsection{Importance of the most forward data points}

Since the data is being extrapolated to zero degrees, the first few data
points
at the smallest angles might hold a special significance. At the same time,
the
most forward data points can be the most poorly determined as they often 
involve a larger background caused, for example, by the beam scattering off the
spectrometer.
To better understand this, we have repeated the above analysis with the
first
two data points having larger errors than the rest. In Table~\ref{table2} we
show results where the errors for the data are set at 1\% with the exception
of
the first two points which have errors of 2\% or 5\%.

Increased errors on the first two data points increases the errors on the
extracted values of the $A$'s noticeably. Roughly, the errors in the
extracted
values of the $A$'s is doubled by having an error of 2\% or 5\% on the first
two data points. This result led us to examine what happens if these first
two
data points were removed from the data set. Does it have a positive effect
to
include very forward data points knowing they are inferior or would it be a
better strategy  to include only the most reliable data points? The results
with the two most forward points removed are given in Table~\ref{table3}.

For $\pi^-$ the removal of the first two data points roughly doubles the
error
for the extracted value of $A_I$ when the errors in the data are 1\%  or
2\%.
For $A_R$ this is also true at $P_{lab}= 500$ Mev/c but at 1 GeV/c  the
increase in the error for $A_R$ is closer to a factor of three to four. As
noted earlier, Re $F_N(0)$ is atypically small for this case and thus the
errors in its determination are of a typical size but are a large fraction
of
the actual value. For $K^+$ a similar result holds. The error in the value
of
$A_I$ is increased by somewhat more than a factor of two. The error in $A_R$
at
500 MeV/c, however, is little affected, while at 1 GeV/c the increase is
roughly a factor of three. If the data has a 5\% error, the errors on the
extracted values of the parameters are sufficiently large that it is
difficult
to see a pattern in the change caused by the removal of the first two
points.

In general, poor quality data for the first several points is slightly
better than
not including the points. In both cases, the first several points
taken with good statistics improves significantly the quality of the
parameters
extracted from the data reducing the errors by a factor of two or more.

\begin{figure}
\caption{
The two pieces of the scattering amplitude, $f_{c,pt}(\theta)$ and
$F_N(\theta)$,
versus $k^2\sin^2\theta/2$ for elastic scattering of $\pi^-$ from $^{12}$C
at
$P_{lab}=$ 500 MeV/c as calculated in the model of
Ref.~\protect\cite{gie88}.
The solid lines are Re $f_{c,pt}(\theta)$ and Re $F_N(\theta)$,
the dashed lines are Im $f_{c,pt}(\theta)$ and Im $F_N(\theta)$. In both
cases
the Coulomb amplitude is the amplitude which is singular at the origin.
The x-axis corresponds to an angular range of 0$^\circ$ to 15$^\circ$.}
\label{fig1}
\end{figure}

\begin{figure}
\caption{ 
The same as Fig.~\protect\ref{fig1} except the pion momentum is 1 GeV/c 
and the x-axis
corresponds to an angular range of 0$^\circ$ to 7.5$^\circ$.}
\label{fig2} 
\end{figure}

\begin{figure} 
\caption{ 
The same as Fig.~\protect\ref{fig1} except the reaction is elastic 
scattering of $K^+$ from 
$^{12}$C at $P_{lab}=$ 500 MeV/c. The x-axis correspond to an angular range 
of 0$^\circ$ to 30$^\circ$}
\label{fig3} 
\end{figure}

\begin{figure}
\caption{
The same as Fig.~\protect\ref{fig3} except the $K^+$ momentum is 1 GeV/c 
and the x-axis 
corresponds to an angular range of 0$^\circ$ to 15$^\circ$.}
\label{fig4}
\end{figure}

\begin{figure}
\caption{ 
The value of $A_R$ found from fitting $d\sigma_N/d\theta$ utilizing the  
expansion
of $F_N(\theta)$, Eq.~\protect\ref{eq:2}, versus the number $N$ of model data 
points
that were fitted. The reaction is $\pi^-$ elastic scattering from $^{12}$C at
$P_{lab}=500$ MeV/c. The data points begin at $\theta_{min}=4^\circ$ and are 
evenly
spaced with $\delta\theta = 0.5^\circ$. The solid curve is for the case where 
two
terms are kept in the Taylor series; the dashed curved is for three terms. 
The
dotted lines represent the exact value of $A_R \pm 1$\%.}
\label{fig5} 
\end{figure}

\begin{figure}
\caption{
The same as Fig.~\protect\ref{fig5} except $A_I$ is presented.}
\label{fig6}
\end{figure}

\begin{figure}
\caption{
The same as Fig.~\protect\ref{fig5} except the reaction is $\pi^-$ incident 
on 
$^{12}$C at $P_{lab}=1$ Gev/c and $\theta_{min}=2^\circ$.}
\label{fig7}
\end{figure}
\begin{figure}
\caption{
The same as Fig.~\protect\ref{fig7} except $A_I$ is presented.}
\label{fig8}
\end{figure}
\begin{figure}
\caption{
The same as Fig.~\protect\ref{fig5} except the reaction is $K^+$ incident on 
$^{12}$C at $P_{lab}= 500$ MeV/c.}

\label{fig9}
\end{figure}

\begin{figure}
\caption{
The same as Fig.~\protect\ref{fig9} except $A_I$ is presented.}
\label{fig10}
\end{figure}

\begin{figure}
\caption{
The same as Fig.~\protect\ref{fig7} except the reaction is $K^+$ incident on
$^{12}$C at $P_{lab}=1$ GeV/c.}
\label{fig11}
\end{figure}

\begin{figure}
\caption{
The same as Fig.~\protect\ref{fig11} except $A_I$ is presented.}
\label{fig12}
\end{figure}

\begin{table}
\caption {The values of the expansion coefficients $A_R$, $A_I$, $B_R$, and
$B_I$
extracted from the model data for $\pi^-$ and $K^+$ scattering from $^{12}$C
at
$P_{lab}=500$ and 1000 MeV/c.The results for model data with statistical
errors
that have a standard deviation of 1\%, 2\%, and 5\% are presented.}
\begin{tabular}{ccccccccc}
meson & $P_{lab}$ (MeV/c) &$\theta_{min}$ & $\theta_{max}$
       & error  & $A_{R}$ (fm) & $A_I$ (fm)& $B_R$ & $B_I$ \\
\hline
$\pi^-$ & 500 & $4^\circ$ & $15^\circ$ & exact & $-2.75$ & 6.37 & $-5.28$ &
   $-30.5$ \\
&&&&1\% & $-2.77 \pm 0.03 $ & $6.39 \pm 0.03 $ &
$ -6.13 \pm 0.99 $ & $-31.6 \pm 1.0 $ \\
&&&&2\% & $-2.74 \pm 0.09 $ & $6.33 \pm 0.06 $ &
$ -2.33 \pm 3.73 $ & $-28.5 \pm 3.1 $ \\
&&&&5\% & $-2.78 \pm 0.20 $ & $6.41 \pm 0.20 $ &
$ -4.85 \pm 8.51 $ & $-32.9 \pm 8.3 $ \\
\hline
$\pi^-$ & 1000 & $2^\circ$ & $7.5^\circ$ &exact & 0.53 & 11.0 & $-31.0$ &
   $-40.8$ \\
&&&& 1\% & $0.52 \pm 0.06 $ & $10.9\pm 0.2$ &
  $-26.5 \pm 9.5$ & $-40.3 \pm 1.9 $ \\
&&&& 2\% & $0.48 \pm 0.13 $ & $10.9\pm 0.2$ &
  $-19.8 \pm 18.3$ & $-40.7 \pm 3.1 $ \\
&&&& 5\% & $0.40 \pm 0.29 $ & $10.9\pm 0.4$ &
  $-22.1 \pm 20.5$ & $-42.1 \pm 5.7 $ \\
\hline
$ K^+$ & 500 & $4^\circ$ & $12.5^\circ$ & exact & $-2.28$ & 2.20 & $7.60$ &
    $-5.46$ \\
&&&&1\% & $-2.35 \pm 0.07 $ & $ 2.54 \pm 0.23 $ &
  $ 16.5 \pm 8.0 $ & $ -1.63 \pm 4.50 $ \\
&&&&2\% & $-2.30 \pm 0.09 $ & $ 2.25 \pm 0.19 $ &
  $ 10.0 \pm 6.7 $ & $ -4.07 \pm 3.90 $ \\
&&&&5\% & $-2.04 \pm 0.32 $ & $2.37 \pm 1.87 $ &
$ -1.5 \pm 7.4 $ & $ -29.9 \pm 31.7 $ \\
\hline
$ K^+$ & 1000 & $2^\circ$ & $7.5^\circ$ &exact& $-2.44$ & 6.15 & 22.6 &
    $-13.7$ \\
&&&&1\% & $-2.42 \pm 0.06 $ & $6.07 \pm 0.22 $ &
  $18.1 \pm 8.6 $ & $-15.5 \pm 1.8 $ \\
&&&&2\% & $-2.44 \pm 0.09 $ & $5.99 \pm 0.44 $ &
  $16.9 \pm 12.5 $ & $-16.0 \pm 2.5 $ \\
&&&& 5\% & $-2.46 \pm 0.31 $ & $6.44 \pm 0.86 $ &
  $33.9 \pm 20.6 $ & $-17.4 \pm 7.1 $ \\
\end{tabular}
\label{table1}
\end{table}

\begin{table}
\caption {The values of the expansion coefficients $A_R$, $A_I$, $B_R$, and
$B_I$
extracted from the model data for $\pi^-$ and $K^+$ scattering from $^{12}$C
at
$P_{lab}=500$ and 1000 MeV/c. The error on the data is 1\% except for the
first two
data points where the errors are 2\% or 5\% as listed under the column
labeled
``error".}
\begin{tabular}{ccccccccc}
meson & $P_{lab}$ (MeV/c) &$\theta_{min}$ & $\theta_{max}$
       & error  & $A_{R}$ (fm) & $A_I$ (fm)& $B_R$ & $B_I$ \\
\hline
$\pi^-$ & 500 & $4^\circ$ & $15^\circ$ & exact & $-2.75$ & 6.37 & $-5.28$ &
   $-30.5$ \\
&&&&2\% & $-2.76 \pm 0.10 $ & $6.38 \pm 0.07 $ &
$ -5.50 \pm 2.62 $ & $-31.0 \pm 3.3 $ \\
&&&&5\% & $-2.71 \pm 0.06 $ & $6.34 \pm 0.08 $ &
$ -3.40 \pm 4.86 $ & $-29.0 \pm 3.8 $ \\
\hline
$\pi^-$ & 1000 & $2^\circ$ & $7.5^\circ$ &exact & 0.53 & 11.0 & $-31.0$ &
   $-40.8$ \\
&&&& 2\% & $0.49 \pm 0.18 $ & $10.9\pm 0.3$ &
  $-26.9 \pm 12.1$ & $-40.8 \pm 3.9 $ \\
&&&& 5\% & $0.42 \pm 0.18 $ & $11.1\pm 0.3$ &
  $-30.8 \pm 13.0$ & $-43.2 \pm 4.6 $ \\
\hline
$ K^+$ & 500 & $4^\circ$ & $12.5^\circ$ & exact & $-2.28$ & 2.20 & $7.60$ &
    $-5.46$ \\
&&&&2\% & $-2.34 \pm 0.07 $ & $ 2.46 \pm 0.40 $ &
  $ 14.3 \pm 10.3 $ & $ -3.89 \pm 3.24 $ \\
&&&&5\% & $-2.33 \pm 0.08 $ & $2.49 \pm 0.33 $ &
$ 14.6 \pm 9.7 $ & $ -3.26 \pm 4.5 $ \\
\hline
$ K^+$ & 1000 & $2^\circ$ & $7.5^\circ$ &exact& $-2.44$ & 6.15 & 22.6 &
    $-13.7$ \\
&&&&2\% & $-2.42 \pm 0.11 $ & $6.15 \pm 0.36 $ &
  $21.5 \pm 9.9 $ & $-15.1 \pm 2.5 $ \\
&&&& 5\% & $-2.41 \pm 0.13 $ & $6.22 \pm 0.43 $ &
  $23.7 \pm 9.0 $ & $-15.0 \pm 2.3 $ \\

\end{tabular}
\label{table2}
\end{table}

\begin{table}
\caption {The same as Table \protect\ref{table1} except the forward most two
data
points have been removed from the analysis.}
\begin{tabular}{ccccccccc}
meson & $P_{lab}$ (MeV/c) &$\theta_{min}$ & $\theta_{max}$
       & error  & $A_{R}$ (fm) & $A_I$ (fm)& $B_R$ & $B_I$ \\
\hline
$\pi^-$ & 500 & $5^\circ$ & $15^\circ$ & exact & $-2.75$ & 6.37 & $-5.28$ &
   $-30.5$ \\
&&&&1\% & $-2.72 \pm 0.07 $ & $6.34 \pm 0.06 $ &
$ -4.02 \pm 2.98 $ & $-29.2 \pm 2.9 $ \\
&&&&2\% & $-2.80 \pm 0.13 $ & $6.37 \pm 0.11 $ &
$ -4.92 \pm 5.09 $ & $-31.3 \pm 5.2 $ \\
&&&&5\% & $-2.59 \pm 0.25 $ & $6.27 \pm 0.18 $ &
$ 2.3 \pm 11.8 $ & $-26.7 \pm 8.8 $ \\
\hline
$\pi^-$ & 1000 & $3^\circ$ & $7.5^\circ$ &exact & 0.53 & 11.0 & $-31.0$ &
   $-40.8$ \\
&&&& 1\% & $0.35 \pm 0.28 $ & $11.1\pm 0.4$ &
  $-27.8 \pm 15.6$ & $-43.5 \pm 7.1 $ \\
&&&& 2\% & $0.29 \pm 0.61 $ & $11.2\pm 0.6$ &
  $-33.9 \pm 17.9$ & $-47.1 \pm 11.6 $ \\
&&&& 5\% & $-0.11 \pm 1.91 $ & $10.9\pm 0.9$ &
  $6.3 \pm 43.8$ & $-44.9 \pm 16.3 $ \\
\hline
$ K^+$ & 500 & $5^\circ$ & $12.5^\circ$ & exact & $-2.28$ & 2.20 & $7.60$ &
    $-5.46$ \\
&&&&1\% & $-2.32 \pm 0.07 $ & $ 2.40 \pm 0.47 $ &
  $ 12.9 \pm 10.8 $ & $ -4.30 \pm 3.21 $ \\
&&&&2\% & $-2.33 \pm 0.11 $ & $ 2.58 \pm 0.45 $ &
  $ 16.3 \pm 12.7 $ & $ -3.58 \pm 5.38 $ \\
&&&&5\% & $-2.19 \pm 0.23 $ & $2.54 \pm 0.84 $ &
$ 6.0 \pm 5.9 $ & $ -14.4 \pm 19.4 $ \\
\hline
$ K^+$ & 1000 & $3^\circ$ & $7.5^\circ$ &exact& $-2.44$ & 6.15 & 22.6 &
    $-13.7$ \\
&&&&1\% & $-2.36 \pm 0.15 $ & $6.34 \pm 0.66 $ &
  $23.9 \pm 15.2 $ & $-17.8 \pm 13.9 $ \\
&&&&2\% & $-2.25 \pm 0.34 $ & $6.61 \pm 1.09 $ &
  $27.4 \pm 20.6 $ & $-22.1 \pm 9.2 $ \\
&&&& 5\% & $-1.82 \pm 1.01 $ & $7.23 \pm 2.10 $ &
  $32.9 \pm 27.1 $ & $-34.7 \pm 26.3 $ \\
\end{tabular}
\label{table3}
\end{table}

\begin{table}
\caption {The values of the expansion coefficients $A_R$, $A_I$, $B_R$, and
$B_I$
extracted from the model data for $\pi^-$ and $K^+$ scattering from $^{12}$C
at
$P_{lab}=500$ and 1000 MeV/c. In order to simulate systematic errors, the
model
data are increased/decreased by 5\% and 10\% as indicated.
}
\begin{tabular}{ccccccccc}
meson & $P_{lab}$ (MeV/c) &$\theta_{min}$ & $\theta_{max}$
       & sys err & $A_{R}$ (fm) & $A_I$ (fm)& $B_R$ & $B_I$ \\
\hline
$\pi^-$ & 500 & $4^\circ$ & $15^\circ$ & $-10$\% & $-2.91$  & 6.17 &
     $-6.95$ & $-34.5$ \\
&&&&$-5$\% & $-2.85$ & 6.28 &
  $-6.22$ & $-32.7$ \\
&&&& exact & $-2.75$ & 6.37 & $-5.28$ &
   $-30.5$ \\
&&&&$+5$\% & $-2.68$ & 6.46 &
  $-3.70$ & $-28.4$ \\
&&&&$+10$\% & $-2.64$  & 6.39 &
     18.7 & $-18.8$ \\
\hline
$\pi^-$ & 1000 & $2^\circ$ & $7.5^\circ$ & $-10$\% & $-0.24$ & 11.0 &
   $-38.2$ & $-49.3 $ \\
&&&& $-5$\% & 0.15 & 11.0 &
   $-35.3$  & $ -44.8 $ \\
&&&& exact & 0.53 & 11.0 & $-31.0$ &
   $-40.8$ \\
&&&& $+5$\% & 0.79 & 10.6 &
   16.1  & $ -42.2 $ \\
&&&& $+10$\% & 1.21 & 10.7 &
   15.1 & $-37.5 $ \\
\hline
$ K^+$ & 500 & $4^\circ$ & $12.5^\circ$ &$-10$\% & $-1.93 $ & 3.66 &
  31.5 & $ -6.91 $ \\
&&&&$-5$\% & $-2.17 $ &  3.11 &
  24.3 & $ -20.4 $ \\
&&&& exact & $-2.28$ & 2.20 & $7.60$ &
    $-5.46$ \\
&&&&$+5$\% & $-2.47 $ &  1.74 &
  6.04 & $ -4.66 $ \\
&&&&$+10$\% & $-2.64 $ & 1.21 &
  5.25 & $ -2.84 $ \\
\hline
$ K^+$ & 1000 & $2^\circ$ & $7.5^\circ$ &$-10$\% & $-1.66 $ & 7.07 &
  37.7 & $-23.7 $ \\
&&&&$-5$\% & $-2.07 $ & 6.66 &
  31.9 & $-17.8 $ \\
&&&& exact & $-2.44$ & 6.15 &
  22.6 & $-13.7$ \\
&&&&$+5$\% & $-2.75 $ & 5.58 &
  7.88 & $-15.8 $ \\
&&&&$+10$\% & $-3.10 $ & 5.33 &
  8.33 & $-14.1 $ \\
\end{tabular}
\label{table4}
\end{table}

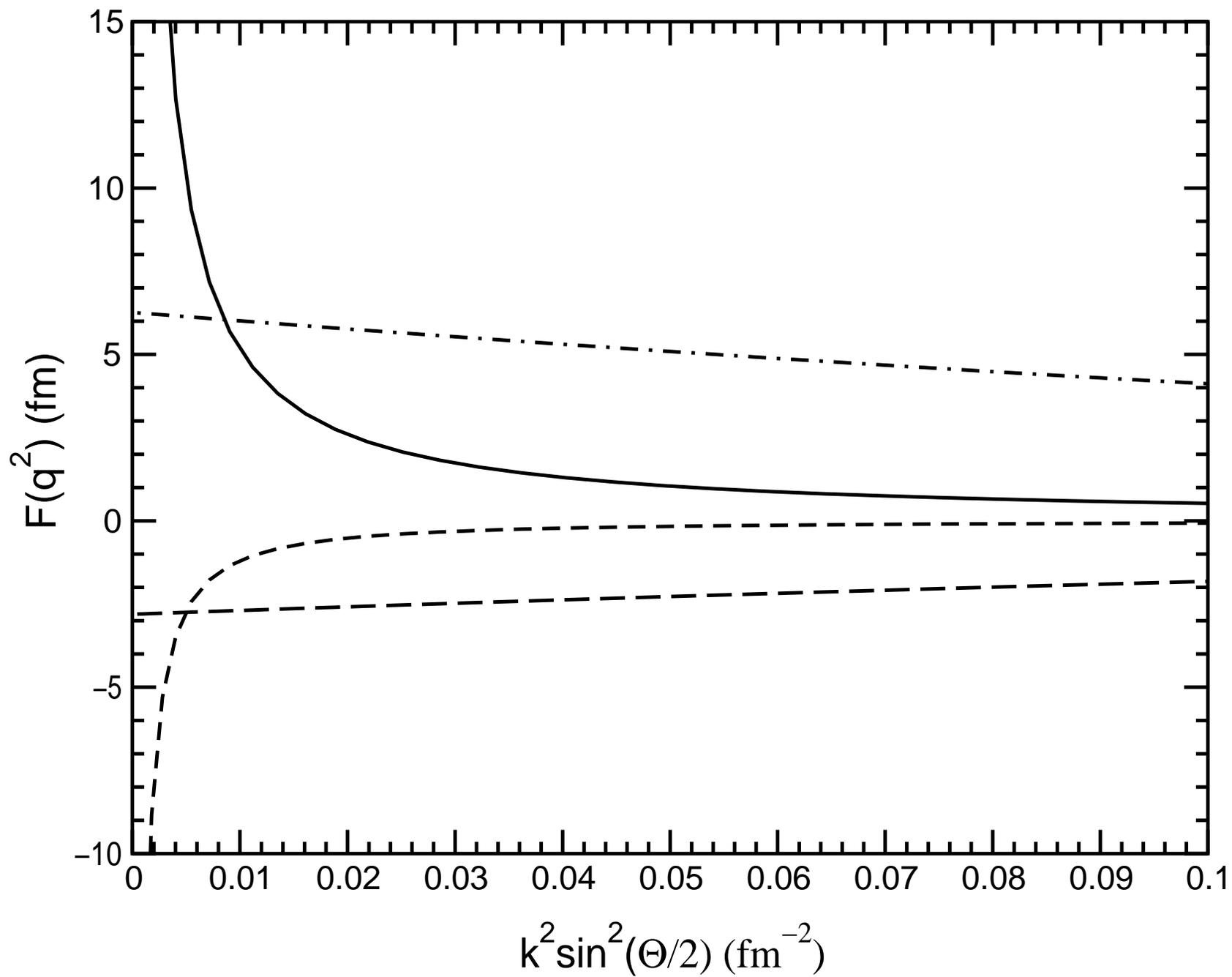
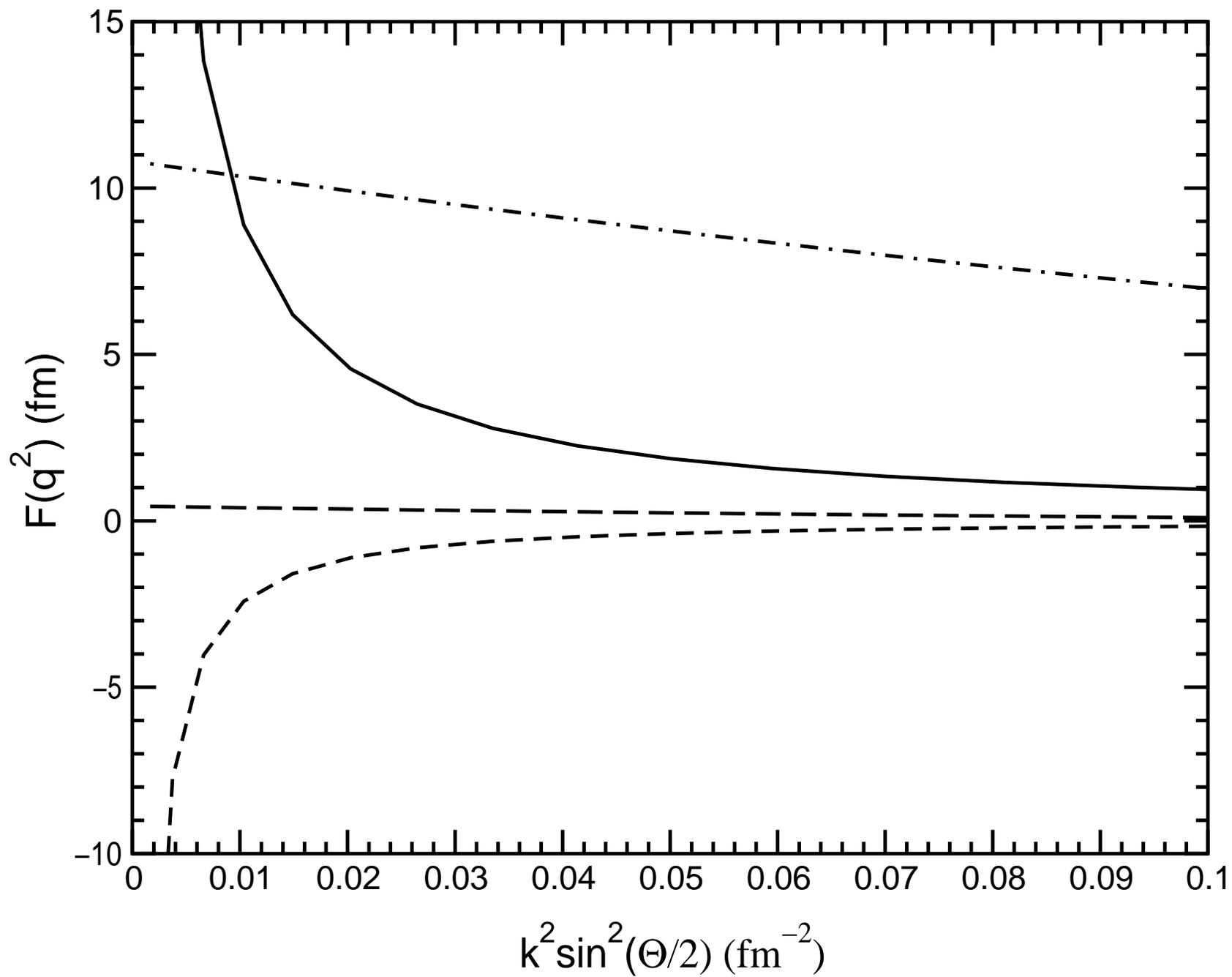
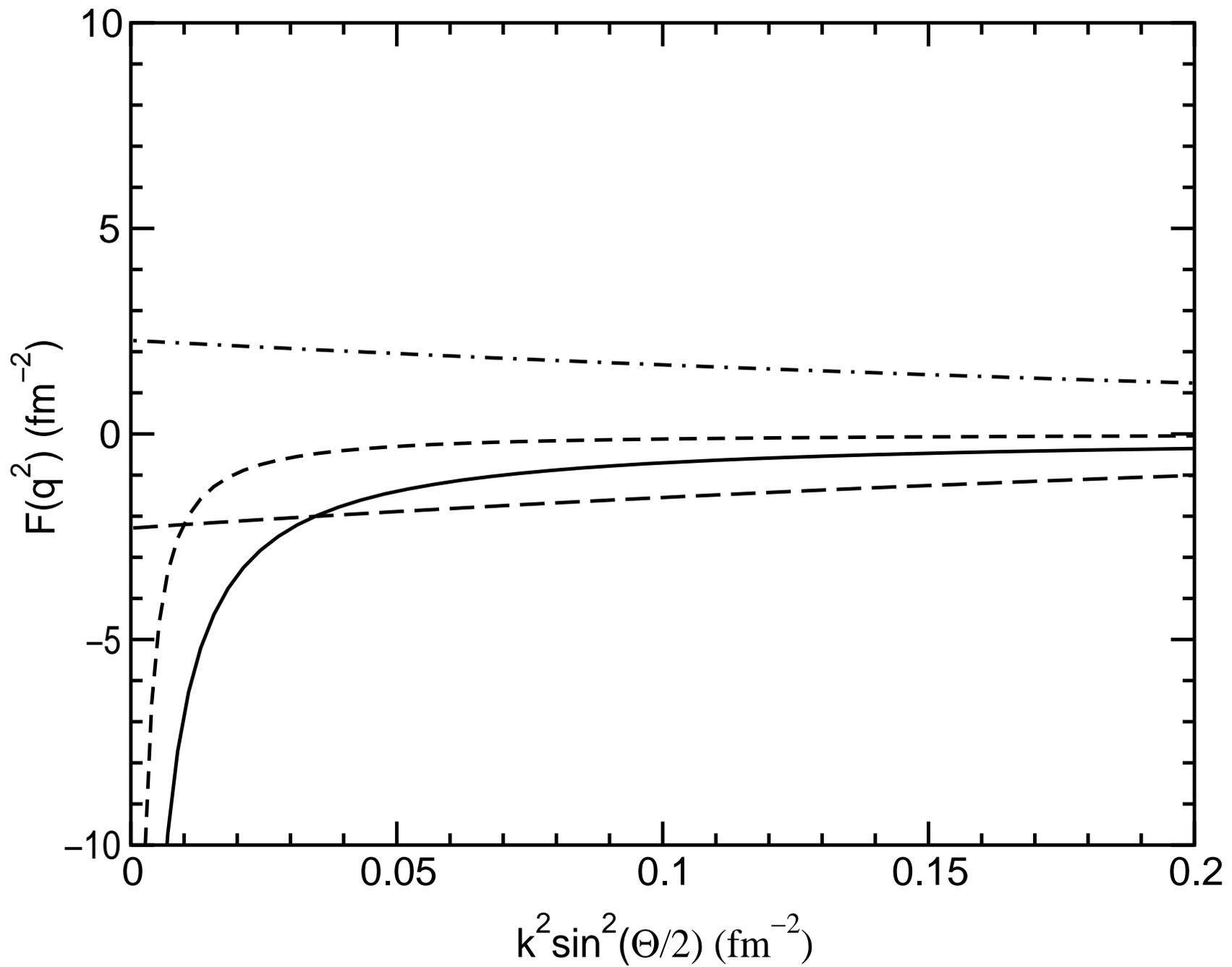
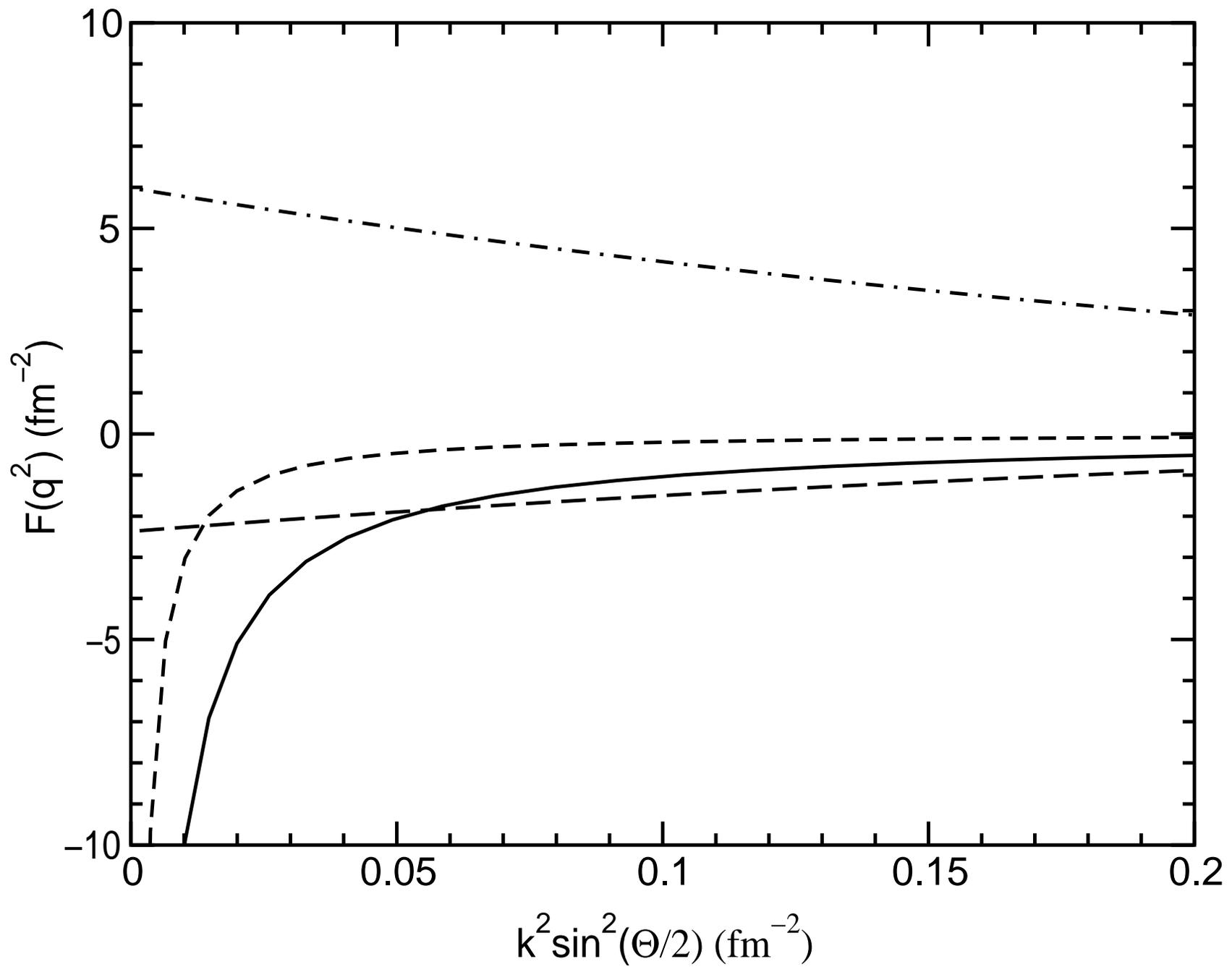
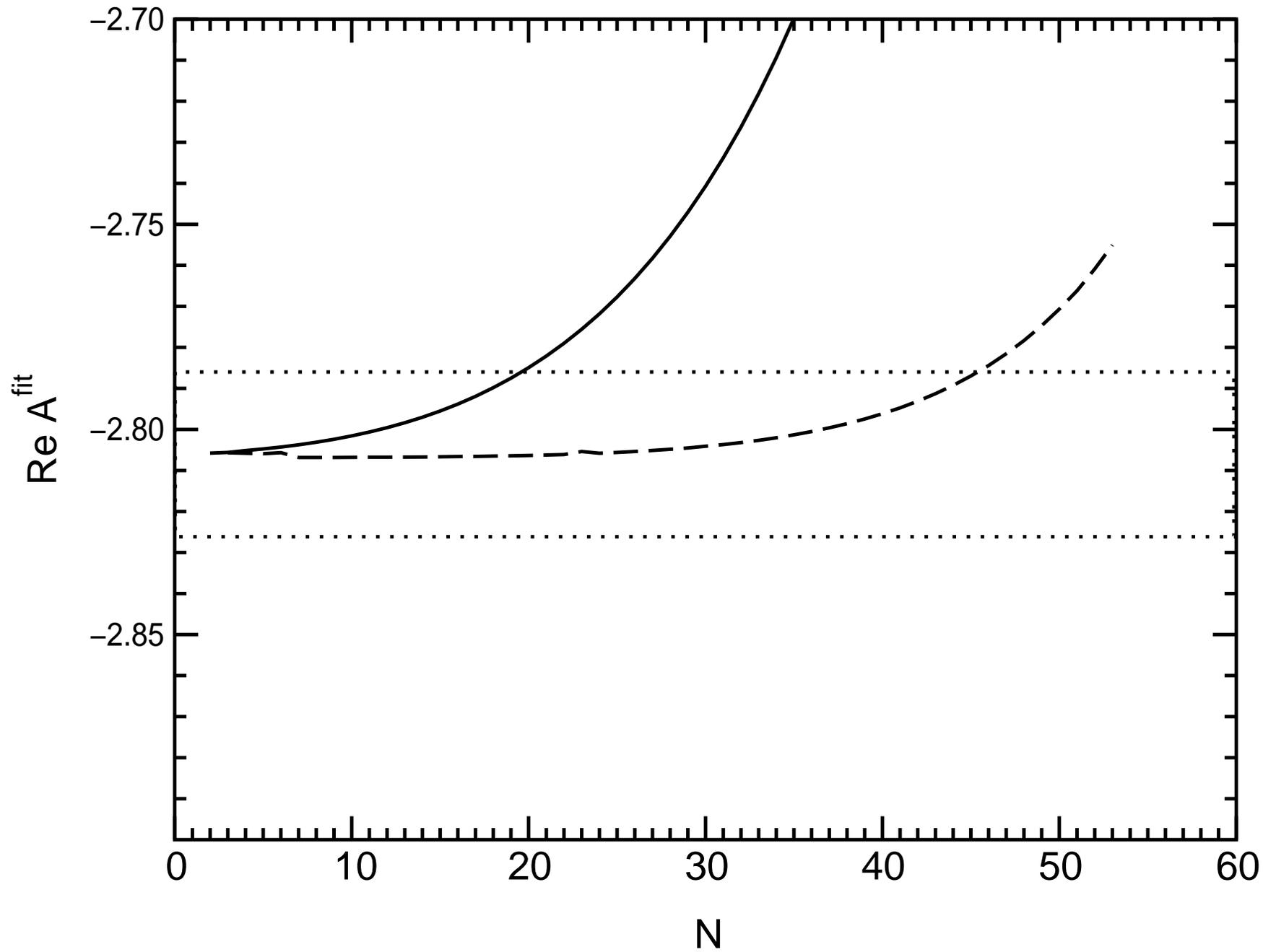
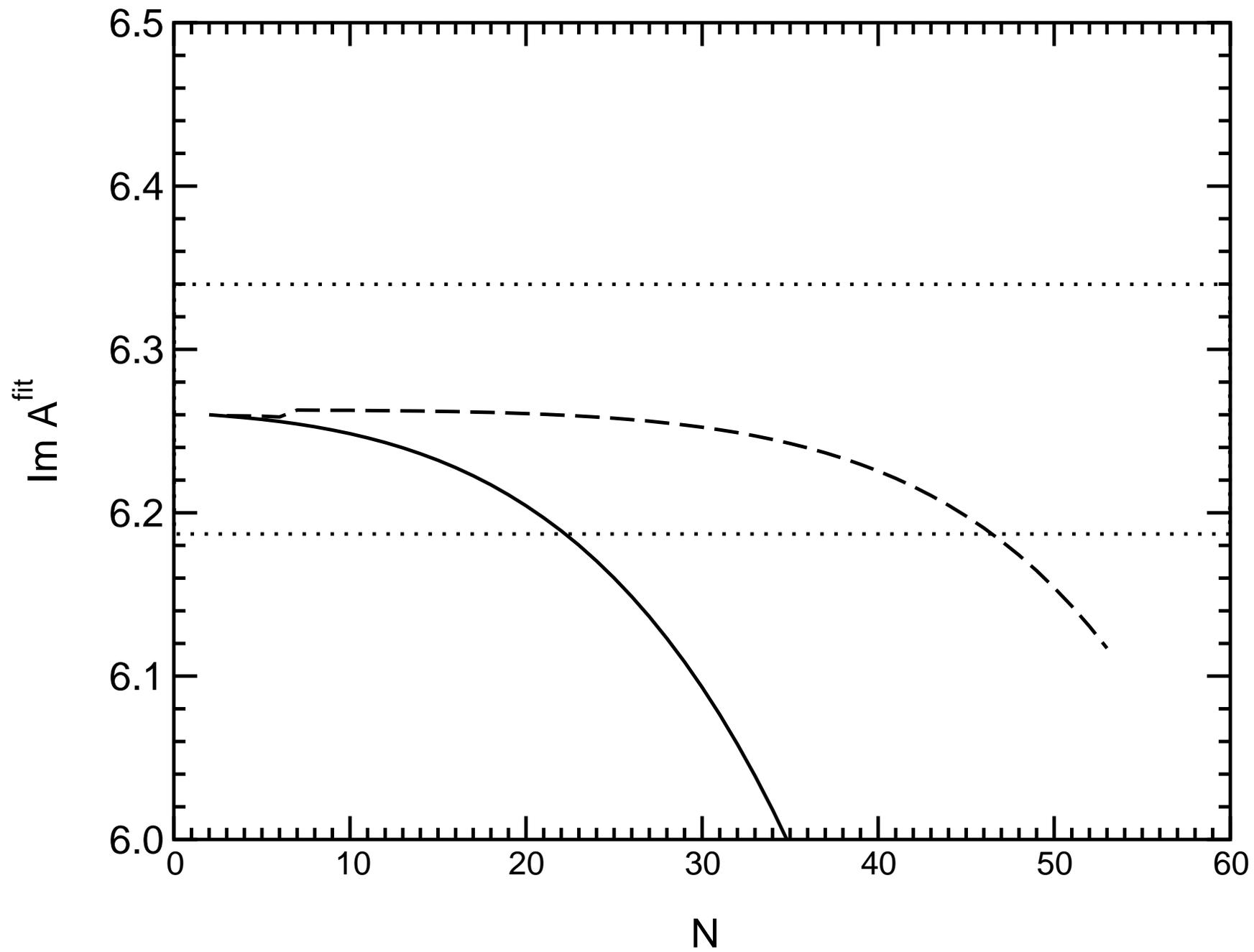
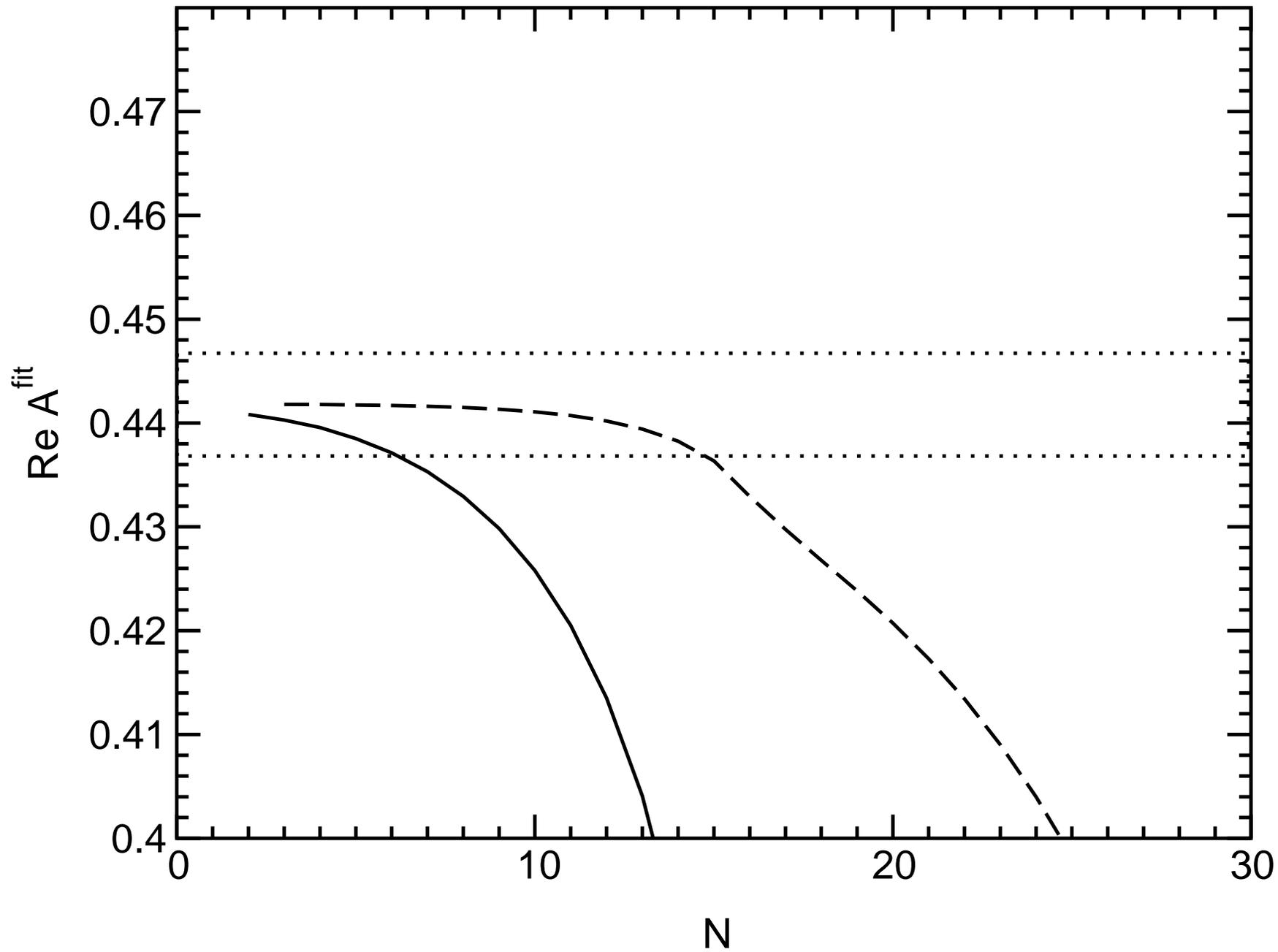
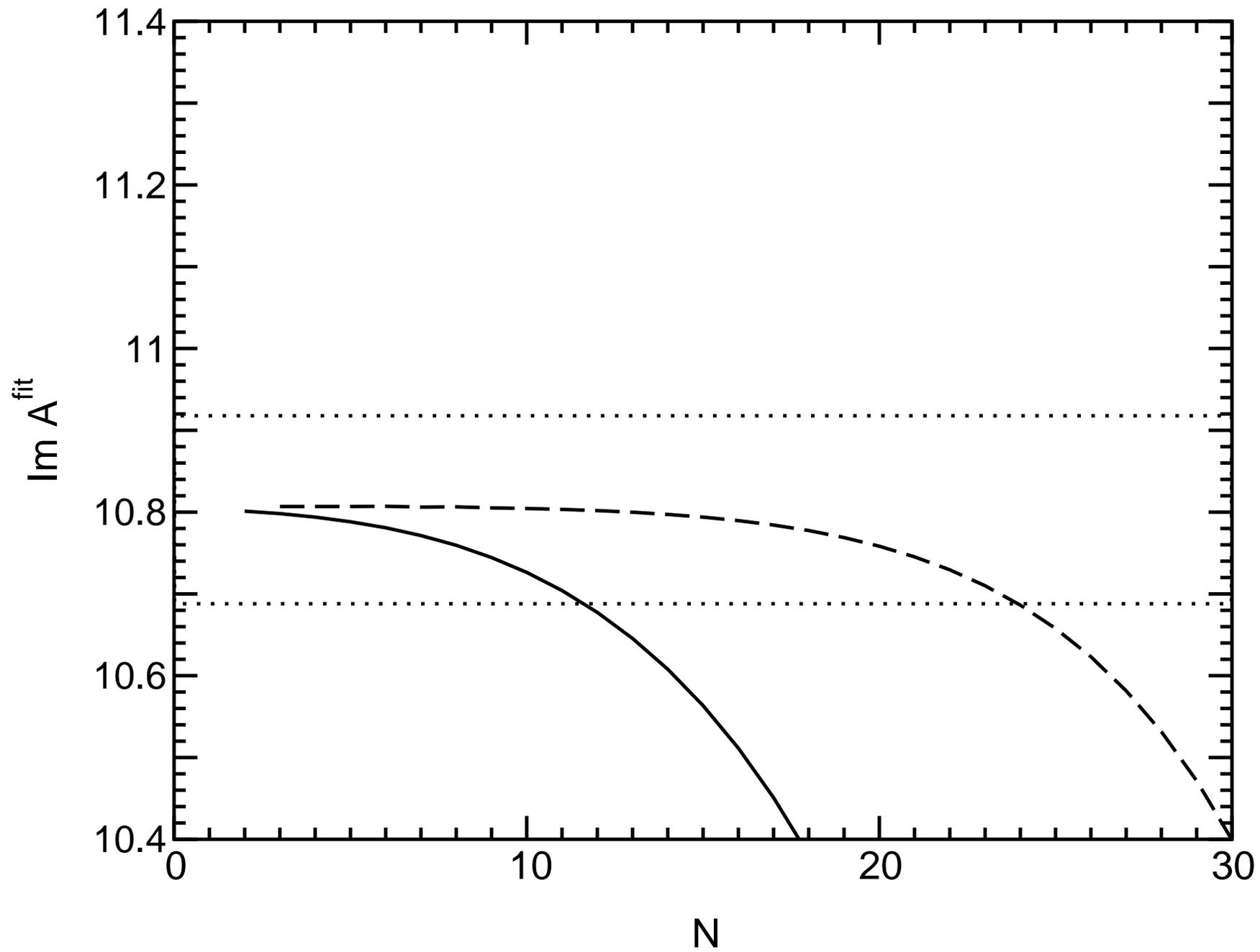
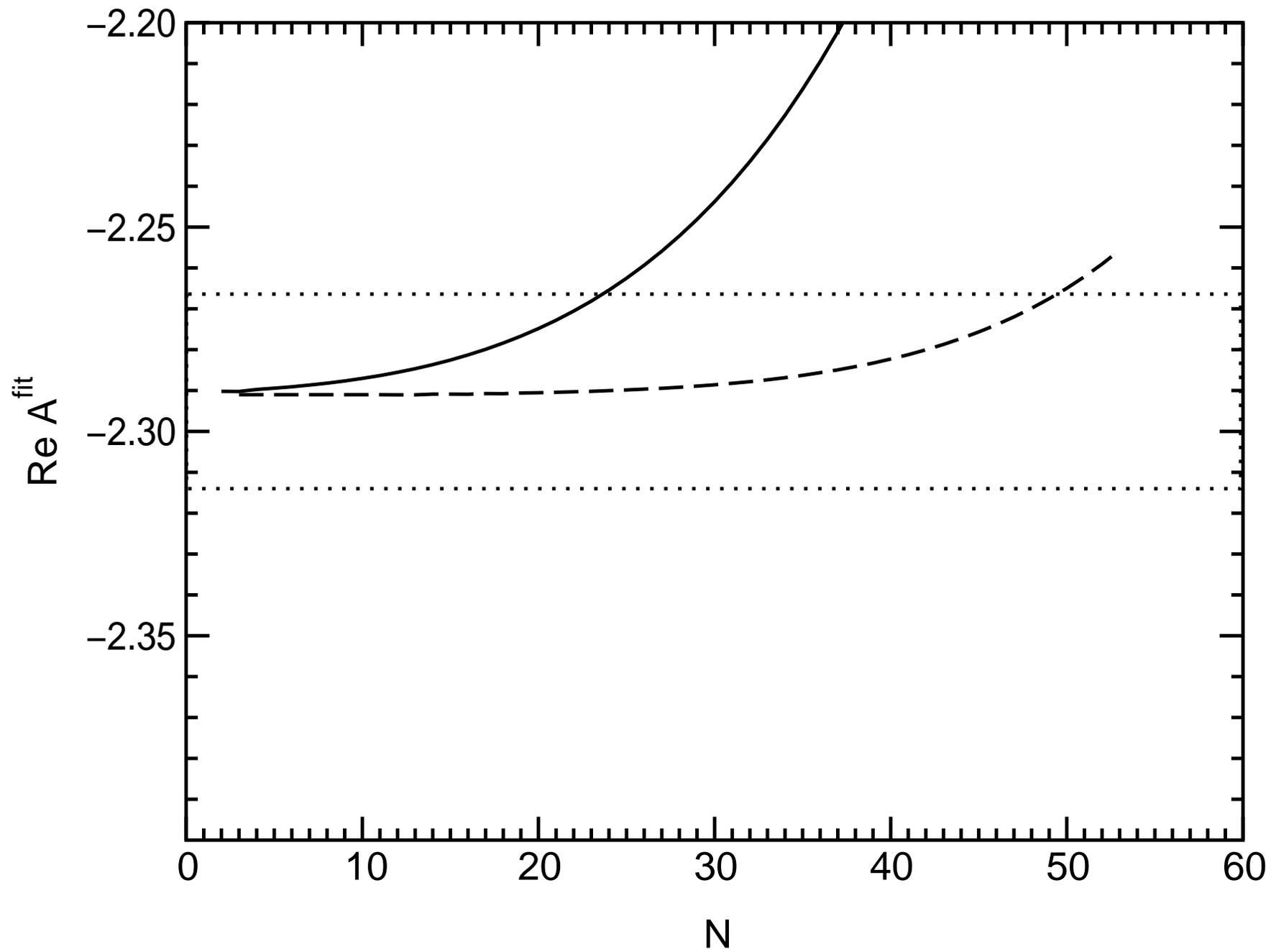
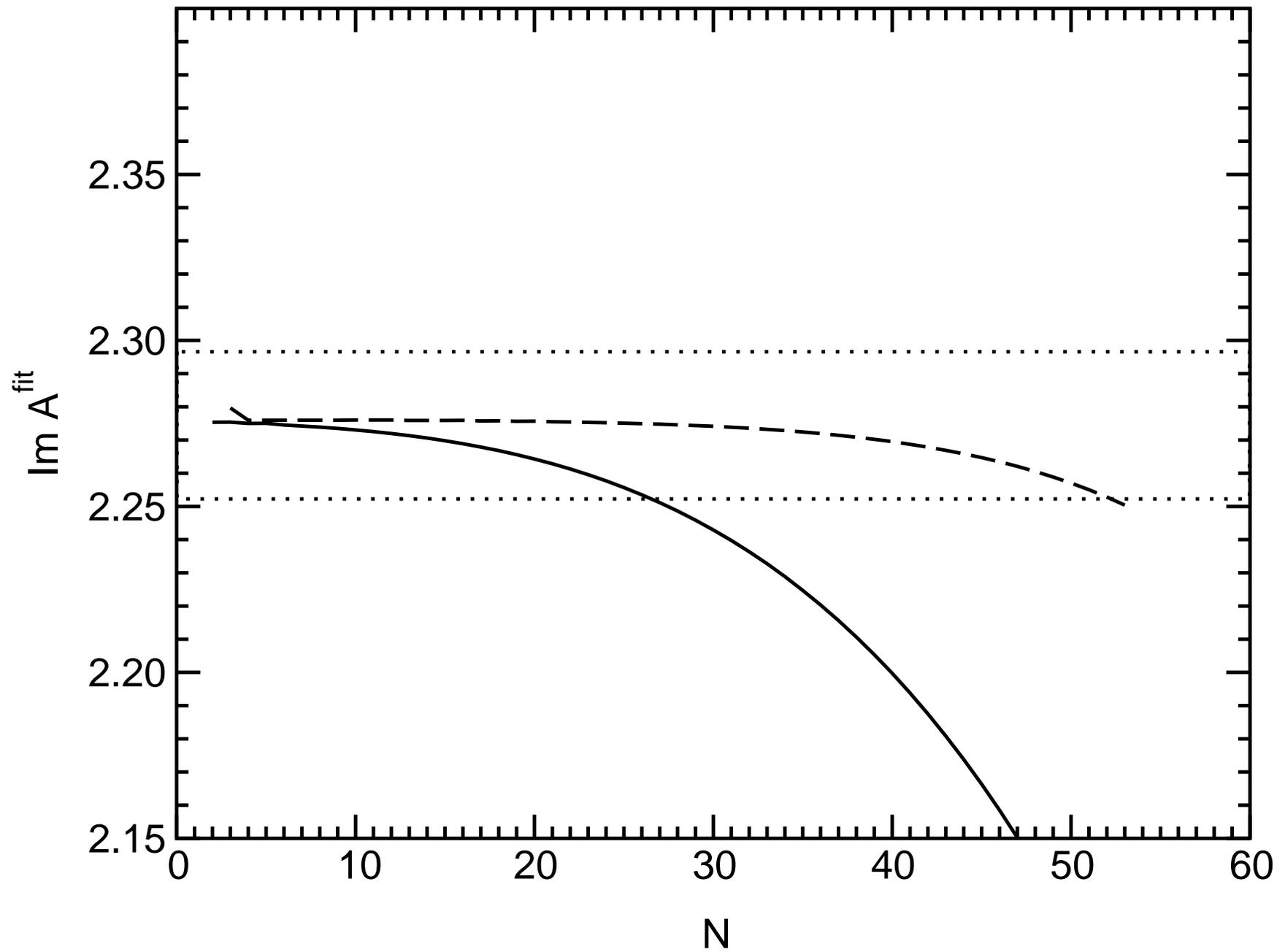
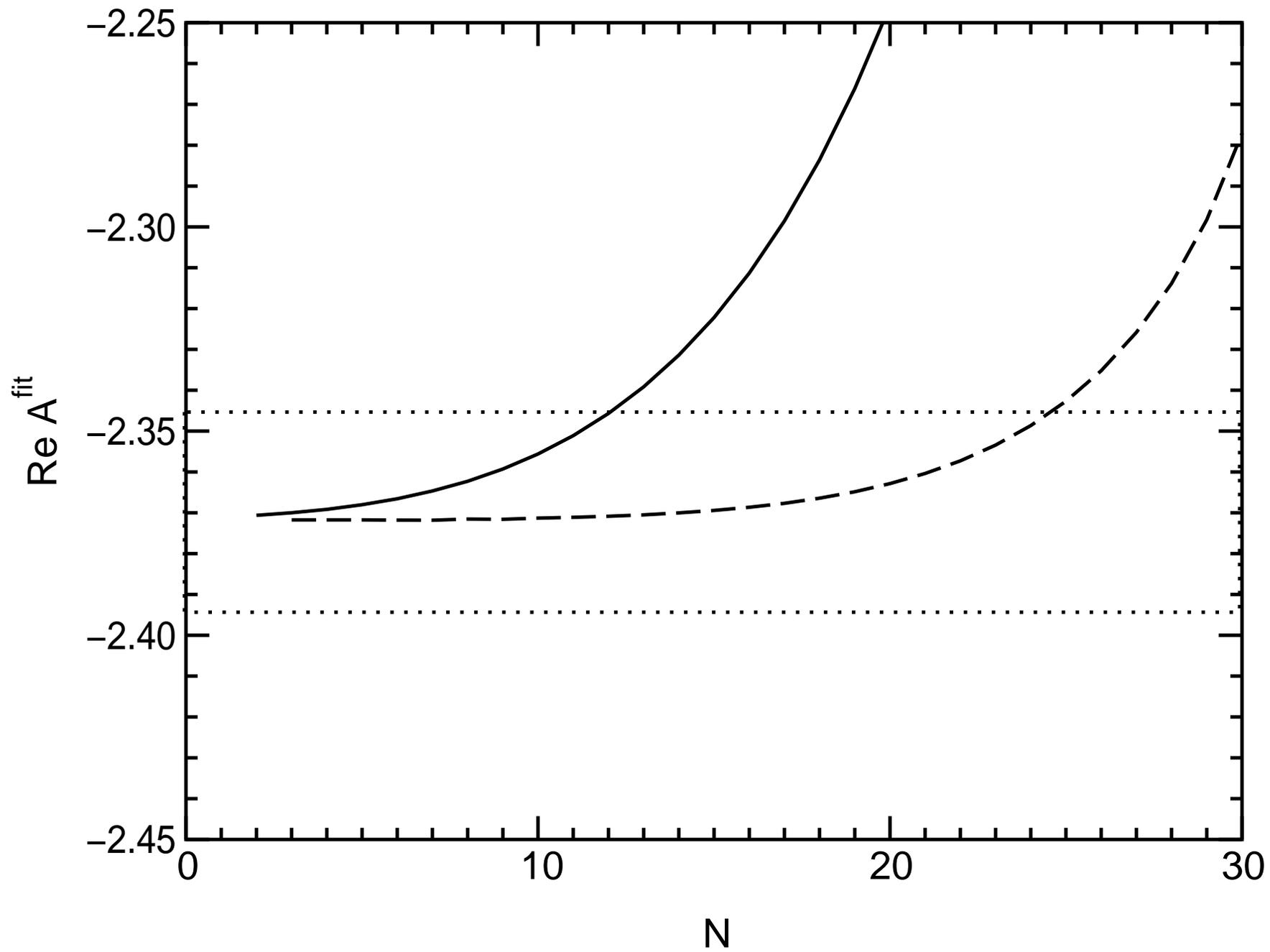
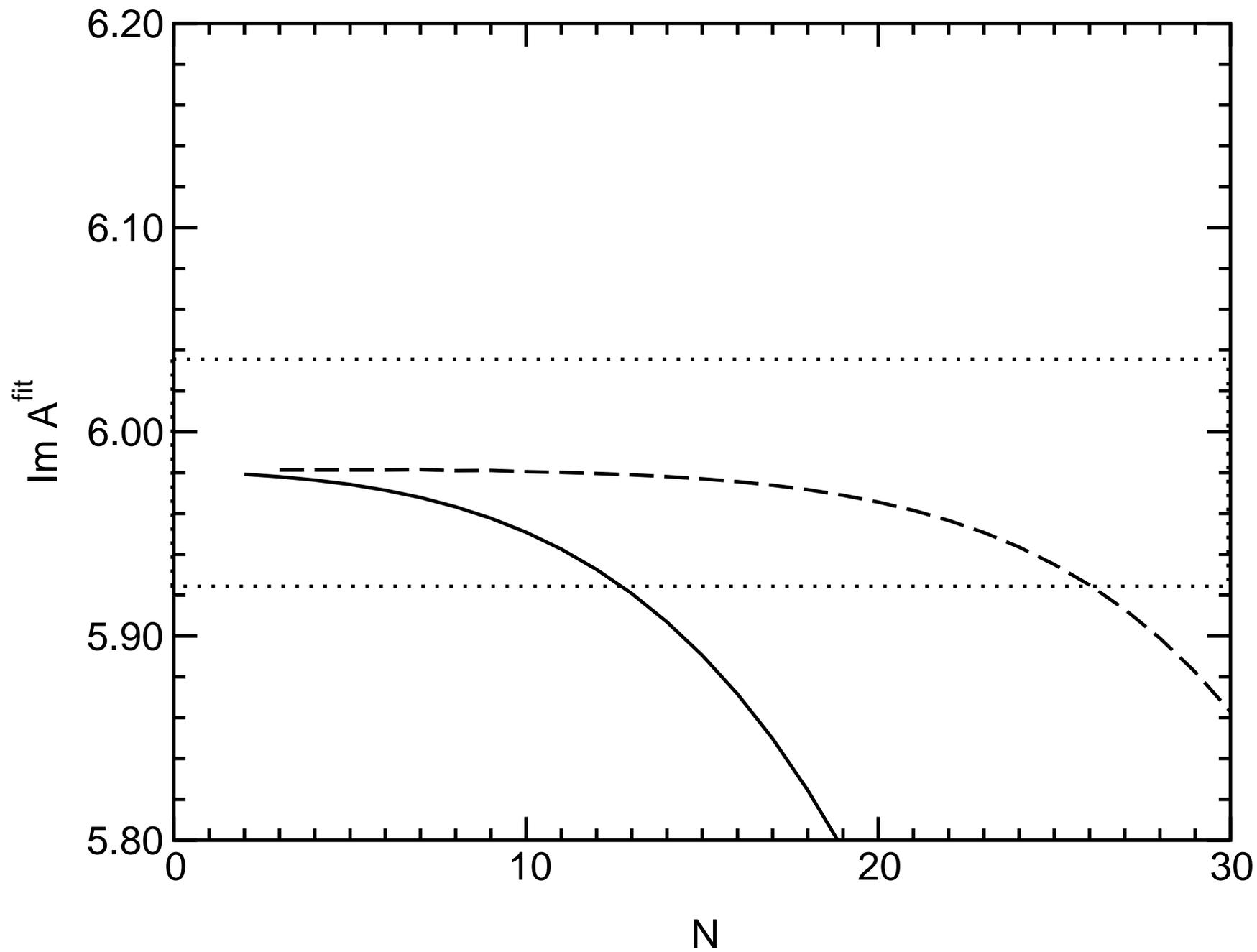
\end{document}